\documentstyle[eqsecnum,aps]{revtex}

\begin{document}
\draft \preprint{HEP/123-qed}
\title{Non Local Electron-Phonon Correlations in a Dispersive Holstein Model}
\author{ Marco Zoli }
\address{Istituto Nazionale Fisica della Materia - Dipartimento di Fisica
\\ Universit\'a di Camerino, 62032, Italy. - marco.zoli@unicam.it}

\date{\today}

\maketitle
\begin{abstract}
Due to the dispersion of optical phonons, long range
electron-phonon correlations renormalize downwards the coupling
strength in the Holstein model. We evaluate the size of this
effect both in a linear chain and in a square lattice for a time
averaged {\it e-ph} potential, where the time variable is
introduced according to the Matsubara formalism. Mapping the
Holstein Hamiltonian onto the time scale we derive the perturbing
source current which appears to be non time retarded. This
property permits to disentangle phonon and electron coordinates in
the general path integral for an electron coupled to dispersive
phonons. While the phonon paths can be integrated out
analytically, the electron path integrations have to be done
numerically. The equilibrium thermodynamic properties of the model
are thus obtained as a function of the electron hopping value and
of the phonon spectrum parameters. We derive the {\it e-ph}
corrections to the phonon free energy and show that its
temperature derivatives do not depend on the {\it e-ph} effective
coupling hence, the Holstein phonon heat capacity is strictly
harmonic. A significant upturn in the low temperature total heat
capacity over $T$ ratio is attributed to the electron hopping
which largely contributes to the action.
\end{abstract}


\narrowtext
\section*{I. Introduction}

Theoretical investigation on the effects of nonlocality in
electron-phonon coupling \cite{sto,per} has grown considerably
during the last years in conjunction with the large interest for
organic molecular crystals \cite{han}, carbon nanotubes
\cite{mah,ver} and conducting polymers exhibiting polaronic
properties. In the Su-Schrieffer-Heeger (SSH) Hamiltonian
\cite{ssh} the electronic hopping is accompanied by a relative
displacement between adjacent atomic sites that causes a nonlocal
{\it e-ph} coupling with vertex function depending both on the
electronic and the phononic wave vector. This feature leads to
peculiar static and dynamical properties \cite{raedt} for the SSH
model which also displays dimension dependent mass behavior for
the acoustical polarons \cite{ono,io1}. In a path integral
investigation of the equilibrium thermodynamics we have shown
\cite{io2} that the purely electronic hopping dominates the low
temperature thermodynamics as it is responsible for a
characteristic upturn in the heat capacity over $T$ ratio pointing
to a glassy like behavior in the one dimensional chain. Moreover,
phonon anharmonicities induced by {\it e-ph} interactions can be
large in the SSH model due to the time retarded nature of the
perturbing source current \cite{io3}.

In the Holstein model, the coupling of electrons to dispersionless
optical phonons is essentially local and the electronic energy at
a lattice site linearly depends on the atomic displacement at that
site \cite{holst}. Thus, in the strong coupling regime, the unit
comprising electron plus induced lattice deformation is generally
a small object (on the lattice scale) undergoing a sizeable mass
renormalization. Large adiabaticities may however induce a spread
in the real space for the Holstein polaron which accordingly
becomes lighter. This effect is more pronounced in higher
dimensionality \cite{atin}. On the other hand, the dispersive
nature of the phonon spectrum is recognized as a fundamental
feature of the Holstein model as it allows for finite values of
the electronic site jump probability while a dispersionless model
would lead to unphysical divergences \cite{holst,prb98}. Also
ground state properties such as electron bandwidth and effective
mass are more properly evaluated versus dimensionality within a
dispersive model. In this paper we investigate to which extent the
realistic assumption of dispersive phonons may induce non local
correlations which consistently renormalize the effective {\it
e-ph} coupling both in one- and two-dimensions. We emphasize that
the model and the results hereafter presented, being valid for any
coupling strength, are independent of the existence of small
polarons in the system.

Unlike the SSH model, the Holstein electron hopping does not
induce a shift in the atomic displacement. Thus the perturbing
{\it e-ph} current does not depend on the electronic paths and,
along the time scale, it turns out to be intrinsecally local.
Applying space-time mapping techniques \cite{hamann} we show that
this feature allows an elegant path integral formulation
\cite{feynman} of the Holstein partition function in which
electron and dispersive phonon coordinates appear to be decoupled.
Section II outlines the Hamiltonian model while the path integral
method is described in Section III. The results of our work are
reported on in Section IV and some final remarks are given in
Section V.

\section*{II. The Dispersive Holstein Model}

We consider the dimension dependent Holstein Hamiltonian
consisting of: i) one electron hopping term, ii) an interaction
which couples the electronic density ($f_{\bf l}^{\dag}f_{\bf l}$)
to the phonon creation ($b_{\bf l}^{\dag}$) and annihilation
($b_{\bf l}$) operators at a given site {\bf l}, iii) dispersive
harmonic optical phonons:

\begin{eqnarray}
& &H =\, H^e + H^{e-ph} + H^{ph} \nonumber
\\
& &H^e =\, - t \sum_{<{\bf l, m}>} f_{\bf l}^{\dag} f_{\bf m}
\nonumber
\\
& &H^{e-ph}=\,  g  \sum_{\bf l} f_{\bf l}^{\dag} f_{\bf l}(b_{\bf
l}^{\dag} + b_{\bf l}) \nonumber
\\
& &H^{ph}=\,  \sum_{\bf q} \omega({\bf q}) b_{\bf q}^{\dag} b_{\bf
q}
\end{eqnarray}

the first sum is over $z$ nearest neighbors, $t$ is the tight
binding overlap integral. $b_{\bf q}^{\dag}$ is the Fourier
transform of $b_{\bf l}^{\dag}$ and $\omega({\bf q})$ is the
frequency of the phonon with vector momentum ${\bf q}$. $g$ is the
{\it e-ph} coupling in energy units.

The phonons operators can be written in terms of the isotropic
displacement field $u_{\bf n}$ as:

\begin{equation}
b_{\bf l}^{\dag} + b_{\bf l} = {1 \over {N}} \sum_{\bf q}
\sqrt{2M\omega({\bf q})} \sum_{\bf n} \exp(i{\bf q} \cdot ({\bf l
- n})) u_{\bf n}
\end{equation}

where $M$ is the atomic mass. Then, the {\it e-ph} term in (1)
transforms as follows:

\begin{equation}
H^{e-ph} =
      {g \over {N}} \sum_{\bf q} \sqrt{2M\omega({\bf q})}
      \sum_{\bf l,n} f_{\bf l}^{\dag} f_{\bf l}u_{\bf n}
      \exp(i{\bf q} \cdot ({\bf l} - {\bf n}))
\end{equation}

The sum over {\bf n} spans all $n^{th}$ neighbors sites of the
{\bf l} lattice site in any dimensionality. Thus, although the
Holstein Hamiltonian assumes a local {\it e-ph} interaction, the
dispersive nature of the phonon spectrum clearly introduces {\it
e-ph} real space correlations which renormalize the dimension
dependent effective coupling. Fourier transforming the atomic
displacement field and taking the lattice constant $|{\bf
a}|=\,1$, from eq.(3) we obtain for a linear chain and a square
lattice respectively:

\begin{eqnarray}
H_{d}^{e-ph} &=&
 {g \over {N^{3/2}}}   \sum_{\bf l} f_{\bf l}^{\dag} f_{\bf l}
      \sum_{\bf q, q^{\prime}} \sqrt{2M\omega({\bf q})}
      \exp(i{\bf q^{\prime}} \cdot {\bf l}) \nonumber
\\{} {} &\times& u_{\bf q^{\prime}}
       S_{d}({\bf q^{\prime} - q})\nonumber
\\
{S}_{1D}({q^{\prime} - q}) &\equiv& 1 + 2
\sum_{n=1}^{n^*}\cos(n(q^{\prime} - q)) \nonumber
\\
S_{2D}({\bf q^{\prime} - q}) &\equiv& 1 + 2 \sum_{n=1}^{n^*}
\Bigl[\cos n(q^{\prime}_x - q_x) + \cos n(q^{\prime}_y -
q_y)\Bigr] \nonumber
\\
{}&+& 2  \sum_{m,n=1}^{n^*} \Bigl[ \cos\Bigl( m(q^{\prime}_x -
q_x) + n( q^{\prime}_y - q_y)\Bigr)   \nonumber
\\ &+& \cos\Bigl(
     m(q^{\prime}_x - q_x) - n(q^{\prime}_y - q_y) \Bigr) \Bigr]
\nonumber
\\
\end{eqnarray}

While, in principle, the sum over $n$ should cover all the $N$
sites in the lattice we introduce the cutoff $n^*$ which will
allow us to monitor the behavior of the coupling as a function of
the range of the {\it e-ph} correlations. Note that in 1D, the
integer $n$ numbers the neighbors shells up to $n^*$ while in 2D,
the $n^*=1$ term includes the second neighbors shell, the sum up
to $n^*=2$ includes the fifth neighbors shells, $n^*=3$ covers the
nineth shell and so on. Switching off the interatomic forces,
$\omega({\bf q})=\,\omega_0$, one would recover from (4) a local
{\it e-ph} coupling model with $S_d \equiv 1$. As no approximation
has been done at this stage eqs.(4) are general.

Taking into account first neighbors intermolecular forces in the
Holstein Molecular Crystal model, the optical phonon spectra are
given in 1D and 2D respectively by \cite{prb98}:

\begin{eqnarray}
\omega^2_{1D}(q)=& &\, {\omega_0^2/2 + \omega_1^2} +  \sqrt {
\omega_0^4/4 + \omega_0^2 \omega_1^2 cosq + \omega_1^2} \nonumber
\\ \omega^2_{2D}({\bf q})=& &\, {\omega_0^2/2 + 2 \omega_1^2}  +
\sqrt { \omega_0^4/4 + \omega_0^2 \omega_1^2  g({\bf q}) +
\omega_1^2 (2 + h({\bf q}))} \nonumber
\\ g({\bf q})=& &\,cosq_x + cosq_y \nonumber
\\  h({\bf q})=& &\, 2cos(q_x -
q_y) \nonumber \\
\end{eqnarray}

where $\omega_0$ and $\omega_1$ are the intra- and inter-molecular
energies respectively.

Treating in (4) the phonon coordinates as classical variables
interacting with quantum mechanical fermion operators we assume a
semiclassical version of the dispersive Holstein Hamiltonian. This
is the main approximation in the model which will permit us to
derive a time dependent source current for the general electron
path integral. Averaging (4) on the electronic ground state we
also define the {\it e-ph} energy per lattice site

\begin{eqnarray}
& &{{<H_{d}^{e-ph}>} \over N} = \sum_{\bf q}<H_{d}^{e-ph}>_{\bf q}
\nonumber
\\ & &<H_{d}^{e-ph}>_{\bf q}=\, {g \over {N^{3/2}}}
\sqrt{2M\omega({\bf q})}  \sum_{\bf q^{\prime}} \rho_{\bf
q^{\prime}} u_{\bf q^{\prime}}
       S_{d}({\bf q^{\prime} - q}) \nonumber \\
& &\rho_{\bf q^{\prime}}=\,{1 \over N} \sum_{\bf l} <f_{\bf
l}^{\dag} f_{\bf l}> \exp(i{\bf q^{\prime}} \cdot {\bf l})
\end{eqnarray}

which will be used in the next Section.

While we are thus neglecting the quantum nature of the lattice
vibrations, the latter may in itself lead to retardation effects
in the ground state structure of the composite quasiparticle made
of one electron plus phonon excitations in the adiabatic regime.
Moreover, at finite temperatures, the quantum lattice fluctuations
are expected to influence the thermodynamics of the system mainly
for intermediate values of the e-ph coupling \cite{raedt1}.

\section*{III. The Path Integral Method}

Let's apply to the Holstein Hamiltonian space-time mapping
techniques \cite{hamann,yuand} which allow us to write the general
path integral for one electron particle in a bath of dispersive
phonons. Thus we introduce ${\bf x}(\tau)$ and ${\bf y}(\tau')$ as
the electron coordinates at the ${\bf l}$ and ${\bf m}$ lattice
sites respectively, and $H^e$ in (1) transforms into

\begin{equation}
H^e(\tau,\tau')=\, -{t} \bigl( f^{\dag}({\bf x}(\tau)) f({\bf
y}(\tau')) + f^{\dag}({\bf y}(\tau')) f({\bf x}(\tau)) \bigr)
\end{equation}

$\tau$ and $\tau'$ are continuous variables $\bigl( \in [0,
\beta]\bigr)$ in the Matsubara Green's functions formalism with
$\beta$ being the inverse temperature hence the electron hops are
not constrained to first neighbors sites. Accordingly, eq.(7) is
more general than $H^e$ in (1). After setting $\tau'=\,0$, ${\bf
y}(0) \equiv 0$, we take the thermal averages for the electron
operators over the ground state of the Hamiltonian thus obtaining
in $d$ dimensions the average energy per lattice site due to
electron hopping:

\begin{eqnarray}
& &h^e(\tau) \equiv {{<H^e(\tau)>} \over N}= \, - {t}\Bigl(G[-{\bf
x}(\tau), -\tau ] + G[{\bf x}(\tau), \tau ]\Bigr) \,\nonumber
\\
& &G[{\bf x}(\tau), \tau]=\,{1 \over {\beta }}\int {{d{\bf
k}}\over {\pi}^d} exp[i{\bf k \cdot x}(\tau)]\sum_n
{{exp(-i\hbar\nu_n \tau)} \over {i\hbar\nu_n - \epsilon_{\bf k}}}
\,\nonumber
\\
\end{eqnarray}

$\nu_n$ are the fermionic Matsubara frequencies and $\epsilon_{\bf
k}=\, -2t\sum_{i=x,y,z}\cos({k_i})$ is the electron dispersion
relation.

The spatial {\it e-ph} correlations contained in (3) are mapped
onto the time axis introducing the $\tau$ dependence in the
displacement field: $u_{\bf q} \to u_{\bf q}(\tau)$. Assuming
periodic atomic particle paths: $u_{\bf q}(\tau + \beta)=\,u_{\bf
q}(\tau)$ we can expand $u_{\bf q}(\tau)$ in $N_F$ Fourier
components:

\begin{equation}
u_{\bf q}(\tau)=\,u_o + \sum_{n=1}^{N_F} 2\Bigl((\Re u_n)_{\bf q}
\cos( \omega_n \tau) - (\Im u_n)_{\bf q} \sin( \omega_n \tau)
\Bigr)\,
\end{equation}

with $\omega_n=\, 2n\pi/\beta$.

Then, on the base of eqs.(6) and (9), we identify the perturbing
source current of the Holstein model with the $\tau$ dependent
averaged {\it e-ph} Hamiltonian term:

\begin{eqnarray}
& &j(\tau)=\, \sum_{\bf q} j_{\bf q}(\tau)\nonumber
\\
& &j_{\bf q}(\tau) = {g \over {N^{3/2}}} \sqrt{2M\omega({\bf q})}
\sum_{\bf q^{\prime}}  u_{\bf q^{\prime}}(\tau) \rho_{\bf
q^{\prime}} S_{d}({\bf q^{\prime} - q})
\end{eqnarray}

With these premises, we are now in the position to write the
general path integral \cite{kleinert} for an Holstein electron in
a bath of dispersive phonons. Assuming a mixed representation, the
electron paths are taken in real space while the phonon paths are
in momentum space. The electron path integral reads:

\begin{eqnarray}
& &<{\bf x}(\beta)|{\bf x}(0)>=\,\prod_{\bf q}<{\bf x}(\beta)|{\bf
x}(0)>_{\bf q}\, \nonumber
\\
& &<{\bf x}(\beta)|{\bf x}(0)>_{\bf q}=\, \int Du_{\bf q}(\tau)
exp\Biggl[- \int_0^{\beta} d\tau {M \over 2} \bigl( \dot{u_{\bf
q}}^2(\tau) \nonumber \\ {} \, &+& \omega^2({\bf q}) u_{\bf
q}^2(\tau) \bigr) \Biggr] \,\nonumber
\\  &\times& \int D{\bf x}(\tau) exp\Biggl[- \int_0^{\beta}d\tau
\biggl({m \over 2} \dot{{\bf x}}^2(\tau) + h^e(\tau) - j_{\bf
q}(\tau)\biggr) \Biggr] \nonumber
\\
\end{eqnarray}

where the kinetic term ($m$ is the electron mass) is normalized by
the functional measure of integration over the electron paths.

As a direct consequence of the time-local nature of the {\it e-ph}
interactions, the Holstein source current does not depend on the
electron path coordinates. Then $j_{\bf q}(\tau)$ in (11) can be
easily integrated over $\tau$ using eqs.(9),(10) and noticing that

\begin{equation}
\int_0^\beta  u_{\bf q}(\tau) = {{ \beta u_o} \over {\sqrt{N}}} {}
\end{equation}

we get:

\begin{eqnarray}
& &\int_0^\beta d\tau j_{\bf q}(\tau)=\,{\beta u_o} {g}_d({\bf q})
\,\nonumber \\ & &{g}_d({\bf q})=\, {g \over {N^2}}
\sqrt{2M\omega({\bf q})} \sum_{\bf q^{\prime}} \rho_{\bf
q^{\prime}} S_{d}({\bf q^{\prime} - q})
\end{eqnarray}

${g}_d({\bf q})$ is thus a time averaged {\it e-ph} potential.

The total partition function can be derived from (11) by imposing
the closure condition both on the phonons (eq.(9)) and on the
electron paths, ${\bf x}(\beta)=\,{\bf x}(0)$. Using (13), we
obtain:

\begin{eqnarray}
& &Z_T=\,\prod_{\bf q} \oint Du_{\bf q}(\tau) exp\Biggl[{\beta
u_o} {g}_d({\bf q}) - \int_0^{\beta} d\tau {M \over 2} \bigl(
\dot{u_{\bf q}}^2(\tau)  \nonumber
\\ &+& \omega^2_{\bf q} u_{\bf q}^2(\tau)
\bigr) \Biggr] \,\times \oint D{\bf x}(\tau) exp\Biggl[-
\int_0^{\beta}d\tau \biggl({m \over 2} \dot{{\bf x}}^2(\tau) +
h^e(\tau) \biggr) \Biggr]  \nonumber
\\
\end{eqnarray}

Eq.(14) shows that the averaged {\it e-ph} coupling is weighed
only by the $\tau$-independent component $u_o$ of the displacement
field. This property will prove to be essential in the next
calculations. The integration over the phonon paths can be done
analytically choosing a measure of integration which normalizes
the kinetic term in the phonon field action:

\begin{eqnarray}
\oint Du_{\bf q}(\tau)\equiv  & & {{\sqrt{2}} \over {(2
\lambda_M)^{(2N_F+1)}}} \int_{-\infty}^{\infty}{du_o} \, \nonumber
\\  \times & &\prod_{n=1}^{N_F} \Bigl( 2\pi n \Bigr)^2
\int_{-\infty}^{\infty} d( \Re u_n)_{\bf q}
\int_{-\infty}^{\infty} d(\Im u_n)_{\bf q} \, \nonumber
\\ \oint Du_{\bf q}(\tau) & & exp\Bigl[-{M \over 2}\int_0^\beta
d\tau \dot{u_{\bf q}}^2(\tau)\Bigr]\equiv 1  \, \nonumber
\\
\end{eqnarray}

being $\lambda_M=\,\sqrt{\pi \hbar^2 \beta/M}$. In the following
calculations we set $M \sim 10^4m$.

Using the result \cite{grad}:

\begin{equation}
\int_{-\infty}^{\infty}du_o \exp(bu_o - cu_o^2) = \sqrt{{\pi}
\over c } \exp\bigl(b^2/4c\bigr)
\end{equation}

we derive:

\begin{eqnarray}
Z_{T}&=&\,\prod_q P({\bf q}) \times \oint D{\bf x}(\tau)
exp\Biggl[- \int_0^{\beta}d\tau \biggl({m \over 2} \dot{\bf
x}^2(\tau) + h^e(\tau) \biggr) \Biggr] \, \nonumber \\ P({\bf
q})&=& \,{1 \over {\beta \omega({\bf q})}}\exp
\Biggl[{{\bigl({g}_d({\bf q}) \lambda_M \bigr)^2} \over {2\pi
\omega({\bf q})^2}} \Biggr] \prod_{n=1}^{N_F} {{(2n\pi)^2} \over
{(2n\pi)^2 + (\beta \omega({\bf q}) )^2}} \, \nonumber \\
\end{eqnarray}

Eq.(17) represents the main analytical result of our model. The
exponential function in $P({\bf q})$ embodies the effect of the
non local correlations due to the dispersive nature of the phonon
spectrum. Phonon and electron contributions to the partition
function are decoupled although the effective potential $g_d({\bf
q})$ carries a dependence on the electron density profile in
momentum space through the function $\rho({\bf q})$.

\section*{IV. Electron-Phonon Correlations and Thermodynamical Results}

First we analyse the behavior of the time (temperature) averaged
{\it e-ph} potential (eq.(13)) in the case of a linear chain and
of a square lattice.

Note that $u_{\bf q}(\tau)$ in (9) is a real quantity consistently
with the closure condition on the lattice displacement path.
Accordingly, from (6) (and (10)), we take a real profile also for
the electron distribution in momentum space. Setting the total
electron density on a single site and defining $\rho_o$ as the on
site electron density, from (6), we write the electron profile
$\rho_{q}=\,\rho_o \cos(q)$ for the 1D system and $\rho_{{\bf
q}}=\,\rho_o \cos(q_x)\cos(q_y)$ for the 2D system. Since the
momentum integration runs over $q_i \in [0,\pi/2]$, $\rho_o$
represents in both cases the total electron density. This choice
is convenient in order to normalize the averaged e-ph couplings
over the same parameter both in 1D and 2D. Physically it
corresponds to pin one electron on a lattice site and to measure
the effects on the potential due to the e-ph correlations with
variable range.

Let's assume low energy phonon spectra parameters,
$\omega_0=20meV$ and $\omega_1=10meV$. Setting $g=\,3$, we take a
strong bare Holstein coupling $g$ although the general trend of
our results holds for any value of $g$. In Fig.1(a), ${{
g}_{1D}({q})/{{\rho_o}}}$ is plotted for three choices of the
cutoff $n^*$ (see eqs.(4)) to emphasize the dependence of the
potential on the {\it e-ph} interaction range. The constant value
of the potential obtained for $\omega_1=0$ is also reported on.
While in the case of short correlations ($n^*=\,4$) there is a
range of wave vectors in which the effective coupling becomes
larger than the dispersionless coupling, long range {\it e-ph}
correlations substantially reduce the effective potential with
respect to the dispersionless case and set in an oscillating
behavior which makes the renormalization q-dependent. The
potential tends to converge for the value $n^*=\,24$ which
corresponds to 48 lattice sites along the chain.

The projections of the two dimensional {\it e-ph} potential along
the $y$ component of the wave vector is reported on in fig.1(b)
for three values of the cutoff on the correlation range. For
$n^*=\,1$ the correlation range is extended to the second neighbor
shell thus including 8 lattice sites. For $n^*=\,2$ and $n^*=\,3$
we normalize over 24 and 48 lattice sites respectively. Then, the
three values of $n^*$ in 2D span as many lattice sites as the
three values of $n^*$ in 1D respectively. This choice permits to
normalize consistently the potential for the linear chain and the
square lattice. There is a strong renormalization for the 2D
effective potential with respect to the dispersionless case for
any value of the wave vector. By extending the correlation range
this tendency becomes more pronounced for $q_y$ close to the
center and to the edge of the reduced Brillouin zone. An analogous
behavior is found by projecting the 2D potential along the $q_x$
axis. The 2D potential stabilizes by including the $9th$ neighbor
shell ($n^*=\,3$) in the correlation range.

Then, an increased range for the {\it e-ph} correlations leads to
an effect which is qualitatively analogous to that one would get
by hardening the phonon spectrum: a reduction of the effective
coupling and the consequent lightening of the quasiparticle
effective mass \cite{alenew}.

Let's see now to which extent the momentum dependent potential
affects the thermodynamics of the system. In the total partition
function (eq.(14)) the electronic and phononic paths have been
integrated out separately as the source current does not depend on
the electronic coordinates. The {\bf q}-phonon contribution to the
partition function is thus given by $P({\bf q})$ in (17) hence,
the free energy due to the {\bf q}-mode is:

\begin{eqnarray}
F_p({\bf q})&=&\, {1 \over \beta}\ln \bigl(\beta \omega({\bf
q})\bigr) - {1 \over \beta}\ln \Bigl( \prod_{n=1}^{N_F}
{{(2n\pi)^2} \over {(2n\pi)^2 + (\beta \omega({\bf q}) )^2}}
\Bigr) \, \nonumber \\ &-& {{ \bigl({g}_d({\bf q}) \hbar \bigr)^2}
\over {2 M \omega({\bf q})^2}}
\end{eqnarray}

where the first two terms represent the harmonic free energy
$F^h({\bf q})$. Since $F^h({\bf q})=\,
\beta^{-1}\ln\bigl[2\sinh\bigl(\beta \omega({\bf
q})/2\bigr)\bigr]$ we find a constraint which allows us to
determine the cutoff $N_F=\,N_F\bigl(\omega({\bf q}),T \bigr)$.
For the 1D system, at $T=\,1K$ and for the phonon parameters given
above, we get $N_F \sim c10^7$ with $c$ varying in the range $[4.2
- 4.5]$ according to the $q$-mode.

The {\it e-ph} interactions renormalize downwards the harmonic
values for any {\bf q} as $F_p({\bf q}) - F^h({\bf q}) \propto -
g^2({\bf q})/\omega^2({\bf q})$ but the temperature derivatives of
the free energy do not involve the coupling term. Then, on general
grounds, phonon entropy and heat capacity are not affected by the
strength of the {\it e-ph} interaction and signatures of {\it
e-ph} anharmonicity should not be expected in the equilibrium
properties of the phonon subsystem. This is a direct macroscopic
effect of the local nature of the interactions in the Holstein
model. On the other hand lattice non linearities may appear in the
ground state properties of the Holstein Hamiltonian and
significantly modify the dynamical properties of the polaronic
quasiparticles \cite{voulga,chris}.

In Fig.2, we consider the linear chain with the cutoff $n^*=\,24$
of Fig.1(a) and set $\rho_o=\,1$. The phonon free energies are
plotted for two selected wave vectors together with the
corresponding harmonic values. It is seen that the renormalization
of the $q$-dependent free energies due to the {\it e-ph} coupling
is very weak. However, being $g^2({\bf q})/\omega^2({\bf q})
\propto g^2/ \omega({\bf q})$, the phonon free energy reduction
may be appreciable in very low energy phonon spectra provided that
the bare coupling $g$ is sufficiently strong.

Next we turn to the evaluation of the electronic contribution to
the total partition function (eq.(17)). The results hereafter
presented refer to the case of 1D electronic paths. As the hopping
energy density is related to the electron propagator (eq.(8)) a
preliminar wave vector integration prepares $h^e$ as a function of
the path coordinates. Then, the $\oint Dx(\tau)$ integration has
to be done numerically. Assuming the periodicity condition
$x(\tau)=\,x(\tau + \beta)$, the particle paths can be expanded in
$N_p$ Fourier components

\begin{eqnarray}
& &x(\tau) \sim \,x_o + \sum_{n=1}^{N_p} 2\Bigl(\Re x_n \cos(
\xi_n \tau) - \Im x_n \sin( \xi_n \tau) \Bigr)\, \nonumber
\\ & &\xi_n=\,2\pi n/\beta
\end{eqnarray}

so that the functional measure of integration

\begin{eqnarray}
& &\oint Dx(\tau) \sim {{\sqrt{2}} \over {(2
\lambda_m)^{(2N_p+1)}}} \int_{-\Lambda}^{\Lambda}{dx_o}
\prod_{n=1}^{N_p}(2\pi n)^2 {\int_{-2\Lambda}^{2\Lambda} d\Re x_n
\int_{-2\Lambda}^{2\Lambda} d\Im x_n} \, \nonumber
\\ & &\lambda_m=\,\sqrt{ \pi \hbar^2 \beta/m}
\end{eqnarray}

normalizes the free electron term in (17):

\begin{equation}
\oint D{\bf x}(\tau) exp\Biggl[- \int_0^{\beta}d\tau {m \over 2}
\dot{\bf x}^2(\tau) \Biggr] =\,1
\end{equation}

Eq.(21) provides a criterion to set the cutoff $\Lambda$ which
turns out to be $\propto \lambda_m$ thus implying that the maximum
amplitude of the particle path coefficients becomes large in the
low temperature range. This is consistent with the physical
expectation that towards high temperatures only a reduced number
of paths contributes to the action which thus tends to the
classical limit. Accordingly, the integration over each Fourier
component in (20) is carried out over a number of points
$N_\Lambda \propto K/\sqrt{T}$ and the electron hopping in the
path integral (18) is weighed over a total number of electron
paths $(N_\Lambda + 1)^{2N_p + 1}$. We find that $K \sim 20$ and
$N_p =\,2$ suffice to ensure numerical convergence. While the free
electron term only scarcely contributes to the electronic action,
the hopping dependent action is dominant also at low temperatures.
Infact, according to our measure of integration, the ensemble of
relevant particle paths (over which the hopping energy density is
evaluated) is $T$ dependent. However, given a single set of path
parameters one can monitor the $h^e$ behavior versus $T$. It turns
out that the hopping decreases by decreasing $T$ but its value
remains appreciable also at low temperatures. Since the $d\tau$
integration range is larger at lower temperatures, the overall
hopping contribution to the total action is relevant also at low
$T$.

The equilibrium thermodynamics of the Holstein model can be
derived from (17). We average the phonon contribution over a set
of 50 modes in the 1D Brillouin zone and take the bare {\it e-ph}
coupling $g=\,3$. Phonon and electron free energies are plotted
separately versus temperature in Fig.3(a). Note that in the low
$T$ range the electron free energy has a positive derivative both
for narrow and wide band values although the range shrinks by
increasing the hopping integral value. For $t=\,0.2eV$, the
electron free energy gets the maximum at $T \sim 60K$ while for
$t=\,4eV$, the maximum occurs at $T \sim 13K$. This feature is
mirrored in the total heat capacity which is reported on in
Fig.3(b) for four $t$ values together with the phonon heat
capacity normalized over the number of modes. As noted above the
latter overlaps the purely harmonic heat capacity no matter how
large the {\it e-ph} coupling may be. Thus the Holstein phonon
heat capacity does not contain signatures of {\it e-ph}
anharmonicity. The total heat capacity over $T$ ratio shows an
upturn at low temperatures as shown in Fig.3(c) due to the
electron hopping mechanism which has already been encountered in
the study of the SSH model \cite{io3}. However, while in the
latter the shape of the low $T$ upturn could be tuned by the
strength of the {\it e-ph} coupling in the source action, in the
Holstein model the low $T$ anomaly has purely electronic origins
as the {\it e-ph} effects are contained in the phonon partition
function. This also explains why the absolute values of the
Holstein heat capacity are much lower than the SSH heat capacity
throughout the whole temperature range.

\section*{V. Conclusions}

In the Holstein model, the {\it e-ph} coupling is local and the
vertex function does not depend on the wave vectors. Nonetheless
non local {\it e-ph} correlations arise in the system once the
dispersive nature of the phonon spectrum is taken into account. I
have studied a semiclassical version of the one electron Holstein
Hamiltonian in which quantum mechanical fermion operators interact
with classical lattice displacements. Applying the path integral
method the real space interactions can be mapped onto the time
scale and, in the Matsubara formalism, we have derived an
analytical expression for the time averaged wave vector dependent
{\it e-ph} coupling. Assuming a force constant intermolecular
model we have evaluated the renormalization of the wave vector
dependent {\it e-ph} coupling, both in one- and two-dimensions, as
a function of the range of the {\it e-ph} correlations. In 1D, for
short range correlations ($n^*=\,4$), I find an enhancement of the
effective coupling with respect to the dispersionless coupling in
a window of phonon wave vectors. By increasing $n^*$, the
effective coupling renormalizes downwards for any $q$ although the
reduction is larger towards the edge of the Brillouin zone. In 2D,
the renormalization is even more pronounced and takes place also
for small values of the cutoff on the correlation range.

Thus, the inclusion of the {\it e-ph} correlations in the
effective coupling has an effect which is similar to that one
would obtain by enhancing the characteristic phonon frequency and
it ultimately leads to reduce the effective mass of the
quasiparticle. The Holstein polaron mass in a dispersive phonon
model is thus lighter than in a dispersionless model.

Mapping the {\it e-ph} interaction onto the time scale we also
find that the perturbing source current of the model depends on
the time only through the atomic displacement field, hence it is
not retarded. Then, the temperature (time) averaged current is
proportional to the averaged atomic displacement and it does not
include the electronic coordinates. This property allows us to
decouple phonon and electron degrees of freedom in the general
path integral and in the total partition function. Such a
disentanglement does not occur in the Su-Schrieffer-Heeger model
\cite{io2} as the source current does depend on the electron path
coordinates. As a physical consequence the renormalized Holstein
{\it e-ph} coupling is frozen in the phonon partition function
while the electron action is dominated by the hopping energy
density. After integrating out analytically the phonon degrees of
freedom we have obtained the {\bf q}-dependent phonon free energy
and estimated the {\it e-ph} corrections (with respect to the
harmonic values) in the thermodynamical properties. It turns out
that the Holstein heat capacity does not show any signature of
phonon anharmonicity induced by the {\it e-ph} interactions. This
marks a striking difference with respect to the thermodynamical
behavior of the previously investigated Su-Schrieffer-Heeger
model. On the other hand, the electron hopping energy strongly
contributes to the total action of the Holstein model and, at low
temperatures, it is responsible for a peculiar upturn in the {\it
heat capacity over temperature} ($C_V/T$) ratio. While
qualitatively the latter feature had also been envisaged in the
SSH heat capacity, in the Holstein model the shape of the broad
upturn has purely electronic origins and it is not affected by the
{\it e-ph} source action. This explains why $C_V/T$ has a negative
temperature derivative throughout the whole $T$ range for any
value of the hopping integral.

\begin{figure} \vspace*{12truecm} \caption{
(a) Temperature averaged {\it e-ph} coupling (in units $meV
\AA^{-1}$ versus wave vector for a linear chain. $n^*$ represents
the cutoff on the {\it e-ph} correlations. $\rho_0$ is the
electron density. $\omega_0=20meV$ and $\omega_1=10meV$. $g=3$.
The dispersionless {\it e-ph} coupling is obtained for
$\omega_0=0$. (b) Temperature averaged {\it e-ph} coupling in two
dimensions versus the $y$ component of the momentum. The cases
$n^*=1$, $n^*=2$ and $n^*=3$ imply that the correlation range
includes the second, the fifth and the nineth neighbors shell,
respectively. The input parameters are as in (a).}
\end{figure}

\begin{figure} \vspace*{8truecm} \caption{
Anharmonic $F_p$ and harmonic $F^h$ phonon free energies for two
values of the one dimensional wave vector $q=\,0, \pi/2$. The
phonon spectrum parameters are as in Fig.1(a). }
\end{figure}

\begin{figure} \vspace*{12truecm} \caption{
(a) Phonon free energy (symbols) and electron free energies for
four values of the hopping integral $t$. The phonon energies and
the {\it e-ph} coupling are as in Figs.1. (b) Total (electron plus
phonon) heat capacities for four values of $t$. The phonon heat
capacity is plotted separately. (c) Total heat capacities over
temperature ratio.}
\end{figure}

\section*{Acknowledgements}

The author acknowledges Prof. A.N.Das for fruitful discussions and
a critical reading of the paper. This work is part of the Joint
Research Project (Ph-T 4) under the Indo-Italian Programme of
Co-operation in Science and Technology 2002-2004. I'm thankful to
Dr. G.Costantini for his effective collaboration.

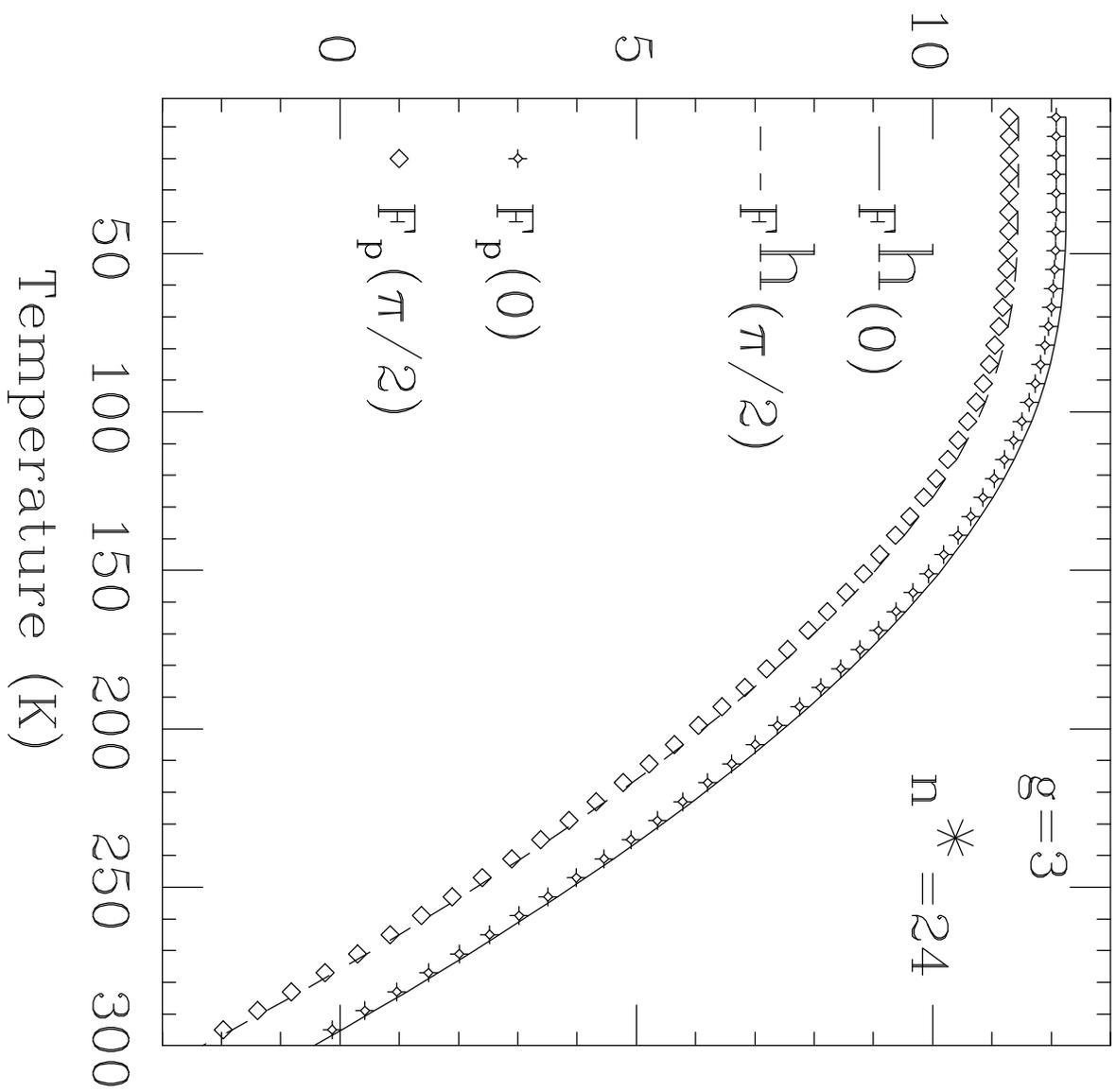
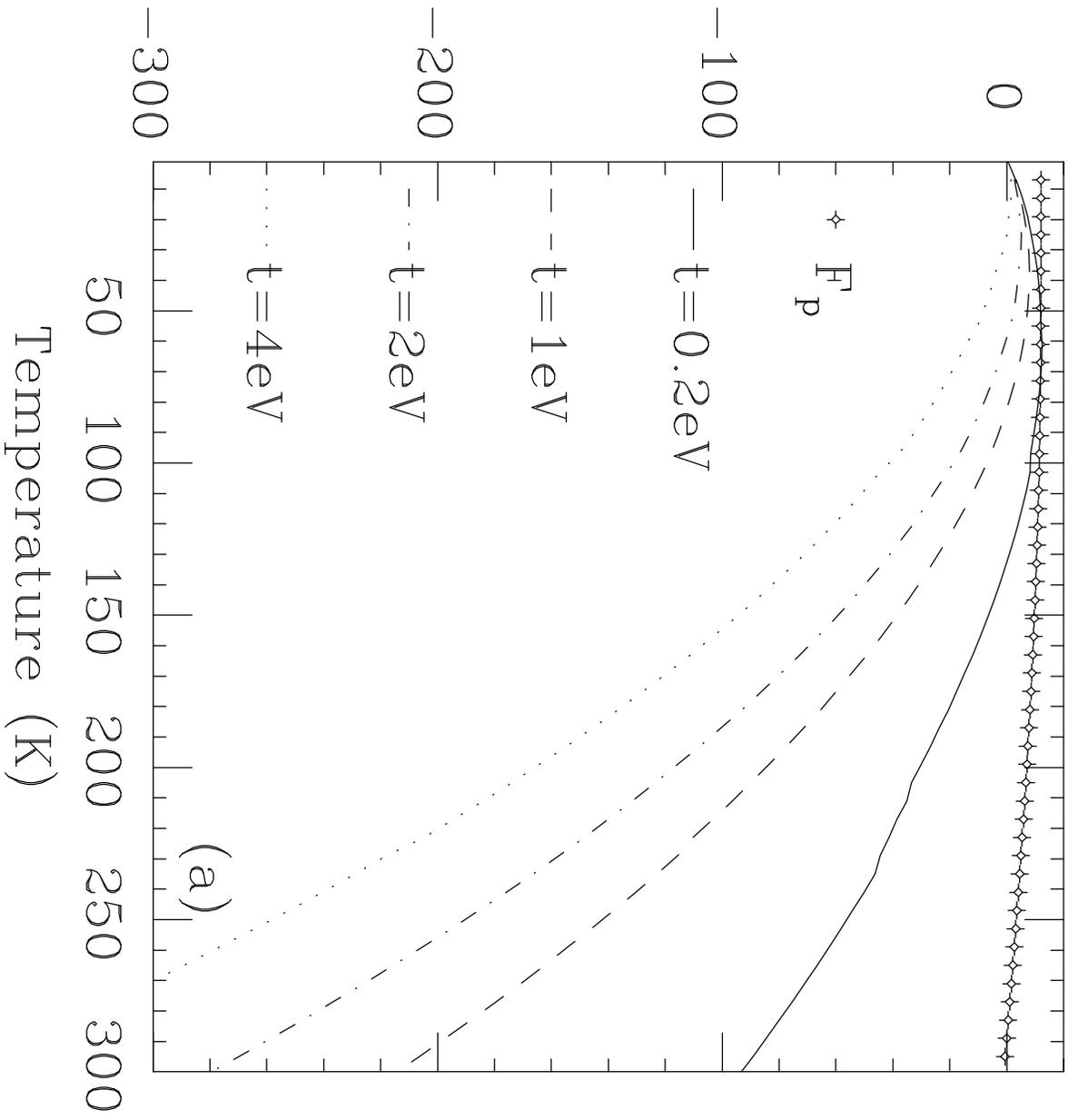
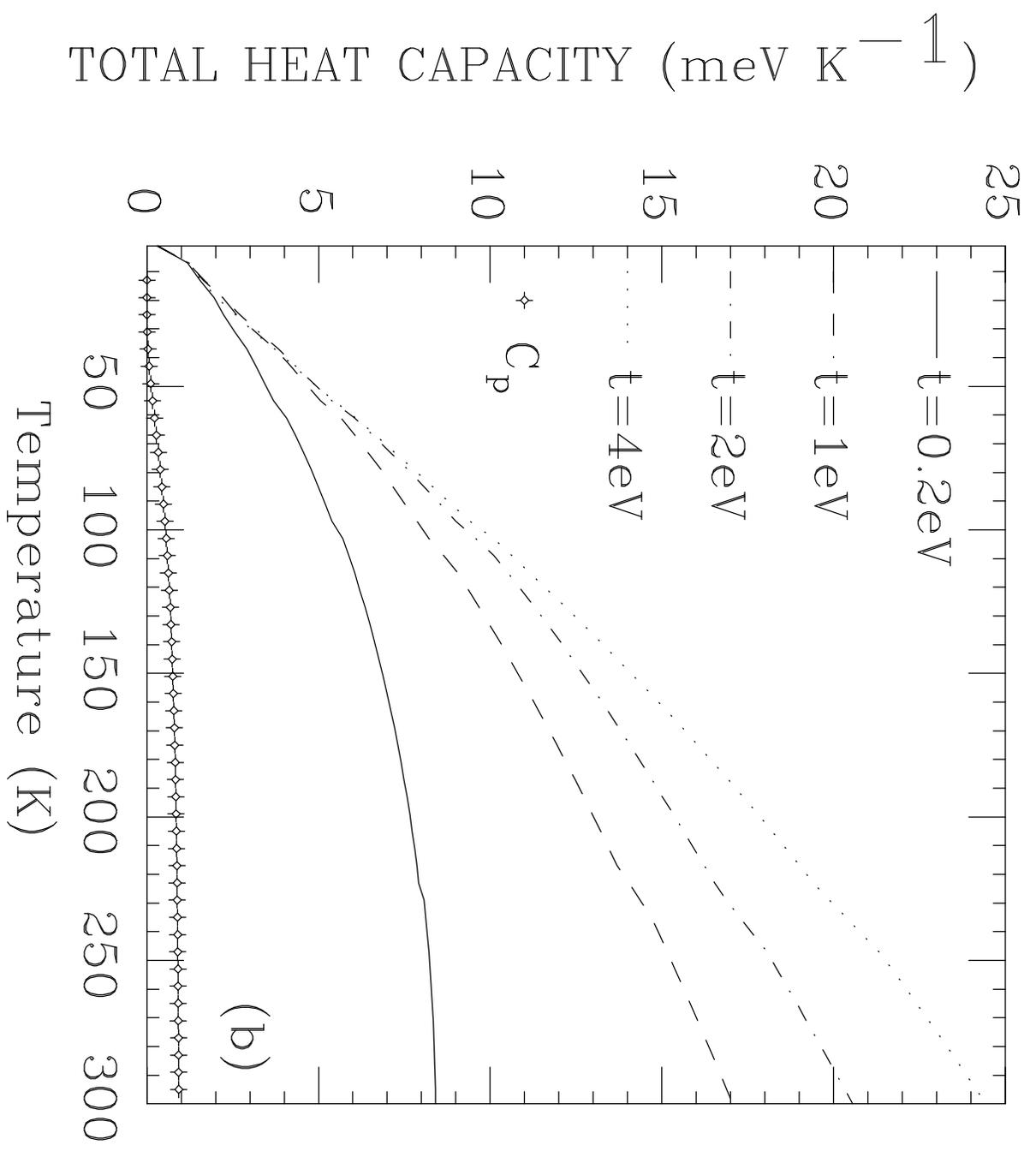
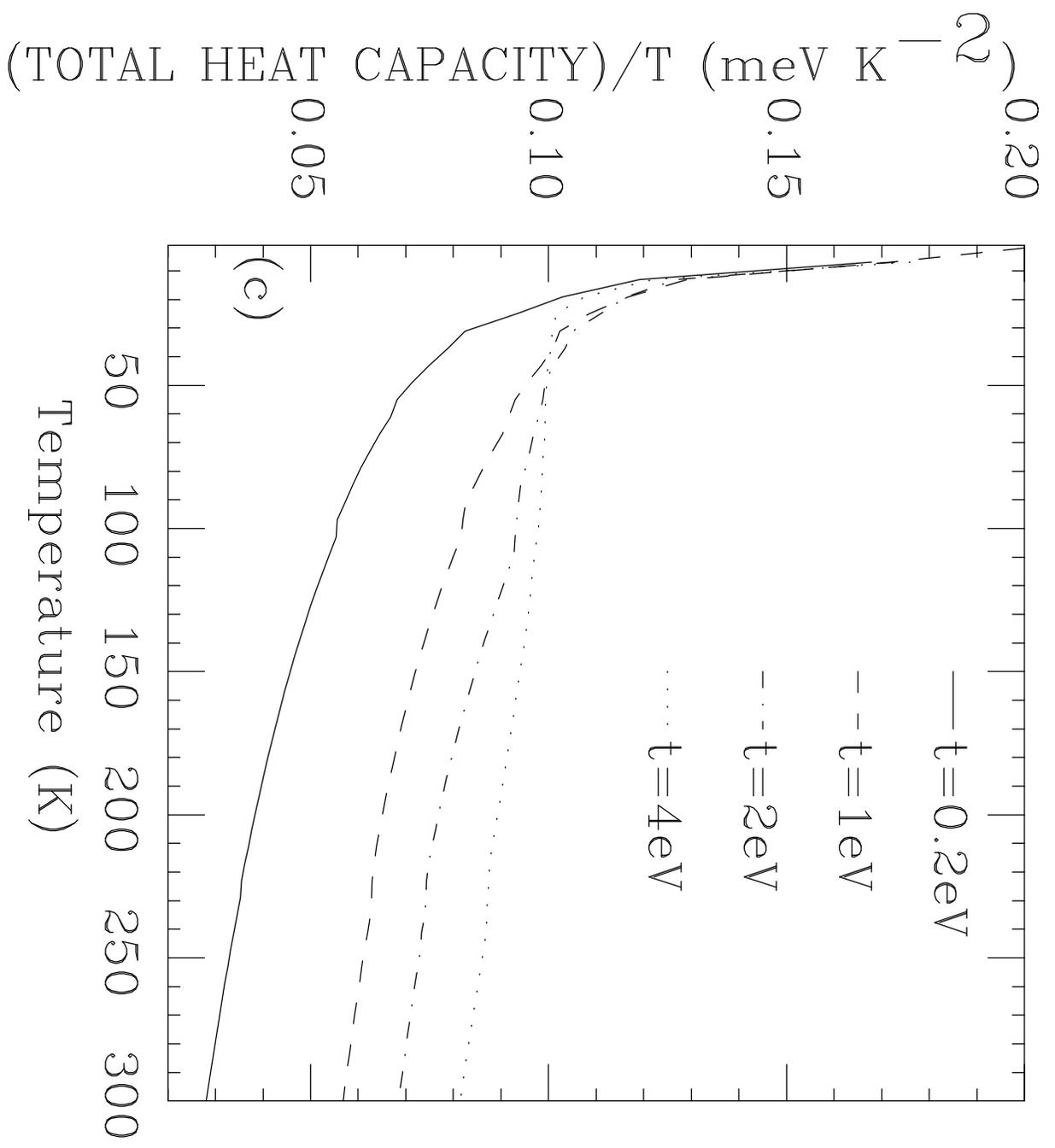
\end{document}